Massimo Cerdonio[1] and Carlo Rovelli[2]
[1]INFN Section and University, via Marzolo 8, I-35131 Padova, Italy,
[2]Aix Marseille Universite, CNRS, CPT, UMR 7332, 13288 Marseille, France
Universite de Toulon, CNRS, CPT, UMR 7332,
83957 La Garde, France
(July 3rd, 2016)

Comment to  "How does Casimir energy fall? IV. Gravitational interaction of regularized quantum vacuum energy" by K A Milton et al., Phys Rev D **89** 064027 (2014)

*Abstract*
The Paper actually concerns a toy model, not physical Casimir cavities made of conducting plates, but the results are taken implicitly to apply in general. We question on general physical grounds one basic assumption and the results of a renormalization procedure. Then, for physical systems,  i) considering condensed matter theory/experiments, we find  strong evidence against the conclusive claims concerning a putative and dominating surface energy present individually on the plates, and ii) we propose two experiments with physical Casimir cavities to show in detail that the results of the renormalization in this case look somewhat paradoxical. In any case the proposed experiments appear to be feasible and thus it could be tested if the putative self-energies of the plates are indeed there in a physical Casimir cavity, or if the toy model of the Paper has by contrast no connection with physical reality. However at the moment the authors are not legitimate to issue as conclusive claims statements like " refute the claim sometimes attributed to Feynman that virtual photons do not gravitate."

cerdonio@pd.infn.it
rovelli@cpt.univ-mrs.fr

The Paper is presented to be conclusive of a long standing effort by the Authors and many others [refs quoted and discussed therein], to answer definitively the question in the title "How does Casimir energy fall" and conclude that one can finally "… refute the claim sometimes attributed to Feynman that virtual photons do not gravitate." However this is actually accomplished using as "toy model, massless scalar fields interacting with semitransparent (δ-function) potentials defining parallel plates, which become Dirichlet plates for strong coupling". So one may wonder if their results may not be connected in any obvious way to proper Casimir cavities, where the relevant field is the electromagnetic one and the plates are conductors, which of course are different conditions both in the field and in the boundaries. But in the last Section the Authors state "These calculations show, quite generally, that the total Casimir energy, including the divergent parts, which renormalize the masses of the plates, possesses the gravitational mass demanded by the equivalence principle. Similar conclusions were drawn by Saharian et al. [35] for the finite interactions between Dirichlet, Neumann, and conducting plates." So the reader is led to take that all the conclusions from the toy model hold for a real, physical Casimir cavity made of conducting parallel plates.

Here we don't enter the issue if Casimir effects provide or not a demonstration that free vacuum and/or virtual photons gravitate and/or if the issue is connected to the Cosmological Constant Problem. However for convenience of the reader and for a minimum of completeness, we mention a few recent papers, in which various contrasting views have been taken [1-5] and give reference to recent Resource Letters [6,7].

Here we contend that the Paper on one side may be based on questionable assumptions, and on the other side gives predictions, which appear simply not there *for a realistic physical Casimir cavity*, and in any case could be tested with experiments, that here we propose. So we contend that the Paper is not actually contributing to the general discussion as in refs [1-7], and *a fortiori* the Authors are not entitled to claim the statement quoted above.

As for the questionable assumptions, one appears to be the idealization of the Casimir cavity as *rigid*, when it is in uniform acceleration or equivalently it is in a uniform gravitational field. This assumption is implicitly maintained in the Paper, which builds on their refs [17, 34]. Rigidity in special and general relativity must be treated with care, because an extended body develops internal stresses [8, 9]. The energy associated with such stresses is missing in the calculations of the Paper. Also the plates of the cavity can be massless in the Paper, as they serve only as a boundary conditions, and one discovers at the end of the Paper that they acquire due mass in a renormalization process. In this respect we considered the *gedanken* experiment given in [4], where we introduced explicitly a spring to fully discuss the equilibrium of the Casimir cavity. In particular we found that the plates *must* have a physical non zero mass. In fact, should it not be so, one would get an absurd even in the limit of increasing the stiffness to a point that the compressional waves in the spring would have velocities close to that of light.

As for predictions, according to the Paper the vacuum would induce a surface energy, which would reside on the plates of Casimir cavities. Such a surface energy i) would be independent of their separation, so that also isolated plates would be endowed with it, ii) would be large in comparison to typical Casimir cavities energies, and peculiarly iii) would be crucial to renormalize the masses of the plates in order to let the system obey to the Equivalence Principle (notice that in the logic of the Paper the masses of the plates can well be zero, as the plates operate only as boundary conditions).

The question is then: is such a surface self-energy on the plates really there? If so, first, as it is said to be much larger than the interaction energy in a Casimir cavity, it should be measurable in a Casimir effect experiment, and second, as it is said to be independent on plate separation, it should be manifest also in energy calculations and experiments with isolated plates. About the first point, take the proposal of ref [10] for an experiment to measure *directly* the Casimir energy, using superconducting plates. The concept relies on the effect of the Casimir energy upon the superconducting transition, implying that the Casimir energy is comparable to the condensation energy in the superconductor. Thus, as the experiment appears to be feasible, an energy much larger than the Casimir should be quite relevant, and either be taken in account in that experiment, or measured in this and in other feasible experiments, as those proposed below.

About the second point, if we consider just one plate by itself, we should find that its cohesion energy should contain that surface energy. But a wealth of *ab initio* quantum mechanical calculations exist, which give the cohesion energies of a piece of metal in agreement with experiments, see for a review [11]. It is interesting to notice, with reference to the above, that they can be accurate to the point to be able to correctly predict superconductivity in metallic phases of Si under high pressure [12] - a remarkable prediction as the energy changes associated with superconductivity are pretty small in comparison to the total cohesion energy. No trace in the theoretical methods and in the experimental results of the self-energies calculated in the Paper.

Consider also another system where there is a delicate balance of cohesion energies and where surface energies are critically relevant: liquid $^4$He films near the superfluid transition, which suffer critical fluctuation induced thinning. This is due to the classic Casimir mechanism, where the fluctuations of the order parameter, which regulates the superfluid fraction of the liquid, take the place of the fluctuations of the vacuum. The experiment is in agreement with theoretical predictions, which simply assume that the order parameter vanishes at both film-vapor and film-substrate interface [13]. Again no trace appears of any analogue of the surface energy in question.

Let us comment now about the *ad hoc* renormalization process, which is invoked to let the plates acquire mass in order to obey to the Equivalence Principle. This looks peculiar, because the mass renormalization given in the Paper is a surface effect and thus one is entitled to expect that it will contribute differently for bodies with different area to volume ratio. In any case the mass-energy of a thin plate would be affected by such a surface self-energy orders of magnitude more than for the mass standard, and the renormalization would be *ad hoc* for each sample. By contrast one would soon be able to measure masses in an absolute way, even no more relating to the mass standard in Paris, but rather expressing the kilogram in terms of Planck's constant [14]. It looks quite difficult to us to find a way out to reconcile such a situation. Also, the above argument appears to be quite general and thus can be applied to the toy model of the Authors just as well as we do in the case of physical Casimir cavities. Thus the experiments proposed below with real materials should have a similar impact on the toy model.

All these considerations invite directed experimental efforts, in particular towards the last question above. This experiment appears conceptually [15] simple: one could vaporize a slab of metallic or semitransparent material, measure the heat of vaporization in a given volume, kept at constant known temperature; then measure the pressure, and get from the law of gases the number of moles of metal present. If the experimental volume is sufficiently large in comparison of the covolume of the atoms, one ends up with an ideal gas of non-interacting atoms at the given temperature. The total mass-energy of the gas is thus the atomic mass times the number of atoms present (plus the mass-energy coming from their kinetic energy, which is easily negligible). Now if one had measured the mass [16] of the slab before vaporization, one can check if it equals the mass of atoms plus the heat of vaporization, or if there is a discrepancy. In addition one could make a significant comparison with the cohesion energy calculated ab-initio, if one uses sufficiently thin slabs, as the volume cohesion energy can be made smaller as the slab gets thinner, while the surface energy can be kept constant. Notice that the concept of such an experiment involves only assuming valid General Relativity, in its weak field approximation, and the classic kinetic theory of ideal gases.

Another experiment is possibly simpler in practice. Take a blob of metal or of semi-transparent material, seal it inside a parallelepiped of transparent refractory material, and measure the mass of the system. Then with, say, laser light vaporize the blob. It will be deposited as a film on the interior wall. If the parallelepiped is thin and the material deposited on the thin lateral walls is burned out with appropriate laser beams, one ends up with a plane parallel Casimir cavity. Now the system mass can be measured and compared with the initial one [16]. As the area to volume ratio between blob and film is very different, if one measures the difference in total mass before and after vaporization, one can see if one gets only the textbook Casimir energy, or if there is an additional - larger - contribution uniquely due to the surface energy of each plate in question, as claimed in the Paper. One could go on and vaporize similarly one of the plates to make the material deposit on the other one, so that now one has only one isolated plate, then do again the mass measurements and check if there is any additional contribution over the initial mass of the blob.

As a final remark we notice that three out of four of the Authors appear now to have second thoughts. In fact we find in the Conclusions of ref [17] the following suggestive statement "However, before we can ascribe a finite self-energy to this configuration, we must recognize that terms in the energy density that grow with the distance into the wall require physical interpretation."

In conclusion it appears that there are strong indications that the surface self-energy, crucially calculated in the Paper, may simply not be there in a physical Casimir cavity, so that the mathematical construction of the Paper may have no connection with physical reality. Therefore the title of the Paper and the quoted claims are misleading to the general reader in their conclusiveness. We hope that our comments here will be taken constructively, may be towards the "physical interpretation" asked for as above, and, in any case, one may always try the experiments proposed in [10] and by us here, to clarify definitively the issue .

Acknowledgements
MC is grateful to Flavio Toigo for helpful discussions, and for calling attention respectively to ref [11] and to ref [13]